\documentclass[submission,copyright,creativecommons]{eptcs}
\providecommand{\event}{SOS 2007} 
\usepackage{breakurl}             

\usepackage{graphicx}

\title{Model-Based Security Testing}
\author{Ina Schieferdecker
\institute{Fraunhofer FOKUS\\ Berlin, Germany}
\institute{Freie Universitaet Berlin\\
Berlin, Germany}
\email{ina.schieferdecker@fokus.fraunhofer.de}
\and
Juergen Grossmann
\institute{Fraunhofer FOKUS\\
Berlin, Germany}
\email{juergen.grossmann@fokus.fraunhofer.de}
\and
Martin Schneider
\institute{Fraunhofer FOKUS\\
Berlin, Germany}
\email{martin.schneider@fokus.fraunhofer.de}
}

\def\titlerunning{Model-Based Security Testing}
\def\authorrunning{I. Schieferdecker, J. Grossmann, M. Schneider}
\begin{document}
\maketitle

\begin{abstract}
Security testing aims at validating software system requirements related to
security properties like confidentiality, integrity, authentication,
authorization, availability, and non-repudiation. Although security testing
techniques are available for many years, there has been little approaches that
allow for specification of test cases at a higher level of abstraction, for
enabling guidance on test identification and specification as well as for
automated test generation.

Model-based security testing (MBST) is a relatively new field and especially dedicated
to the systematic and efficient specification and documentation of security test
objectives, security test cases and test suites, as well as to their automated or
semi-automated generation. In particular, the combination of security modelling
and test generation approaches is still a challenge in research and of
high interest for industrial applications. MBST includes
e.g. security functional testing, model-based fuzzing, risk- and threat-oriented
testing, and the usage of security test patterns. This paper provides a survey on
MBST techniques and the related models as well as samples of new methods and tools that are under development in the
European ITEA2-project DIAMONDS.
\end{abstract}

\section{Introduction}

The times of rather static communication in strictly controlled, closed networks
for limited purposes are over, while the adoption of the Internet and other
communication technologies in almost all domestic, economic and social sectors
with new approaches for rather dynamic and open networked environments
overwhelmingly progresses. Social networks, networked communities, cloud
computing, Web X.0, mashups or business process design are just some of the
trends that reflect the tendency towards permanent connections and permanent data
collection. Today's networked systems face the challenge of various security
threats, which is usually met by various protection systems against attacks from the
direct system users. In an interconnected world, software with vulnerabilities
presents a threat not only to individuals but also to companies and public
organizations, and last but not latest to national and international
cooperation. Compared with functional hazards, which tend to be straightforward
and accidental, security threats are often intentional and more persistent. Here
are some facts that highlight the special nature of security threats:

\begin{itemize}
\item Attacks are frequently carried out by well organized groups with a
  commercial background (spamming, extortion, industrial espionage)
\item Multi-stage attacks skilfully combine vulnerabilities on system level and
  organizational level
\item information security risk analysis does often not hold for the complete life time
  of a product (context of product usage may change, new vulnerabilities are
  detected)
\end{itemize}

Practical security involves a sufficient understanding of risk in order to properly address it, manage it and, with care, to eliminate it. National and
international standardization committees provide significant efforts on security evaluation and assessments. It includes classical concepts from security
evaluation using common criteria (CCRA), but also European activities from ETSI
addressing risk and threat analysis (TVRA).

The main aim of information security methods and techniques is the reduction or
elimination of unwanted incidents that may harm the technical infrastructure or,
even worth, its environment. The field of information security has been growing
and reaches far further than simply using cryptography. The purpose of today's
security standards is to specify countermeasures that can protect a system
against certain forms of exploitation. Countermeasures depend on an
understanding primarily on finding and assessing the vulnerabilities that exist
within a system. Testing can be seen as an action to proactively detect such
vulnerabilities. The Software Engineering Institute, USA, highlighted in 2009:
"The security of a software-intensive system is directly related to the quality
of its software."  About 90 percent of all software security incidents are
caused by attackers who exploit known vulnerabilities. Moreover most known
vulnerabilities originate from software faults or design flaws. However, not
every fault or design flaw constitutes vulnerability. Still, systematic testing increases the likelihood of identifying faults and
vulnerabilities during the design, development or setup time of  systems and enables purposeful fixes.

\section{Model-based security testing and security testing }

Software testing is an experimental approach of validating and verifying that a
software system meets its functional and extra-functional requirements and works as expected. In this article, testing refers to active, dynamic testing, where the behavior of a system under test (SUT) is checked by applying intrusive tests that stimulate the system and observe and evaluate the system reactions. This is
done by applying specification-based and/or code-based test cases that are -- based on test hypotheses -- directed to find faults
in the SUT and by providing test suites, i.e. specifically selected
collections of test cases, which provide an argument for the absence of faults.

Software security testing is a special kind of testing with the aim in validating and
verifying that a software system meets its security requirements. Two principal approaches can be used: functional security testing and security vulnerability testing~\cite{GuTian-yang2010}. While security
functional testing is used to check the functionality, efficiency and
availability of the designed and developed security functionalities and/or security systems (e.g. firewalls, authentication and authorization subsystems, access
control), security vulnerability (or penetration) testing directly addresses the
identification and discovery of yet unknown system vulnerabilities
that are introduced by security design flaws or by software defects. Security vulnerability testing uses the simulation of attacks and other
kinds of penetration attempts. 

The systematic identification and reduction of
security-critical software vulnerabilities and of defects will increase the overall dependability of
software-based systems and helps providing adequate security levels for open
systems and environments. Unfortunately security testing, especially security vulnerability testing, lacks
systematic approaches, which enable the efficient and goal-oriented
identification, selection and execution of test cases. Risk-based testing~\cite{Gerrard:2002:RBE:560726} is a methodology that makes software risks the
guiding factor to solve decision problems in the design, selection and
prioritization of test cases. 

\subsection{Related work}

The basic idea of model-based testing (MBT) is that instead of creating test cases manually, selected
algorithms are generating them automatically from a (set of) model(s) of the
system under test or of its environment. While test automation replaces manual test
execution by automated test scripts, model-based testing replaces manual test
designs by automated test generation. Although there are a
number of research papers addressing model-based security (see e.g.~\cite{Basin:2006:MDS:1125808.1125810, juerjens05} and model-based testing (see
e.g.~\cite{Baker2007}), there is until today little work on model-based security
testing (MBST). Relevant publications in the field of MBST are~\cite{RefWorks:887,
Jurjens:2008:MST:1467086.1467133,
  DBLP:conf/ershov/JurjensW01, MouelhiFBT08, Wang:2007, Wimmel02,Kaksonen2001,takanen2008fuzzing}.

Kaksonen et al.~\cite {Kaksonen2001} from the PROTOS project (1999-2001)
discuss and implement an MBST approach using syntax
testing as the starting point, and implement the models using Augmented
Backus-Naur Form (ABNF). The PROTOS approach to model-based testing reads in
context-free grammars (defined by BNF, ASN.1 (Abstract Syntax Notation One),or XML (Extended Markup Language)) for critical protocol interfaces
and generates the tests by systematically walking through the protocol
behaviour. These two approaches are the only ones commercially
available~\cite{takanen2008fuzzing}. 

Jurjens and Wimmel~\cite{DBLP:conf/ershov/JurjensW01, Wimmel02} address the problem of generating
test sequences from abstract system specifications in order to detect possible
vulnerabilities in security-critical systems. Both papers assume that the system
specification, from which tests are generated, is formally defined in the
language Focus -- a mathematical framework for the specification, refinement, and verification of
distributed, reactive systems. The paper~\cite{DBLP:conf/ershov/JurjensW01} focuses on testing
of firewalls whereas~\cite{Wimmel02} focuses on transaction systems. In~\cite{Jurjens:2008:MST:1467086.1467133}, Jurjens extends~\cite{Jur02} by
considering system specification written in the language UMLsec -- a security profile for the Unified Modelling Language. In~\cite{RefWorks:887}, Blackburn et. al. summarizes the results of applying a
model-based approach to automate functional security testing. The approach
involves developing models of security requirements as the basis for automatic
test vector and test driver generation. In particular, security requirements are
written in the so-called SCRtool, are transformed into test specifications which
in turn are transformed into test vectors and test drivers. The approach is
targeted towards Java applications and database servers. 

Mouelhi
et al.~\cite{MouelhiFBT08} propose a model-driven approach to specifying,
deploying and testing access-control policies in Java applications. The approach
has four main steps. The first step is to build a platform-independent
access-control model for the application. In the second step, the model is
transformed into so-called platform specific policy decisions points (PDPs). In
step three, the PDP is integrated into the functional code of the application by
aspect oriented programming techniques. Finally, in step four, the resulting
integrated application is tested against tests that are generated from the
platform independent access-control model. Another approach covering
specification, deployment, testing and monitoring of security policies has been
proposed in the Politess project (Grenoble INP, IT, Smartesting) [DFGMR06,
MBC08, LR07].  In~\cite{Wang:2007}, Wang et. al. presents a threat driven
approach to MBST. In this approach, UML sequence
diagrams to specify threat a model, i.e., event sequences that should not occur
during the system execution. The threat model is then used as a basis for code
instrumentation. Finally, the instrumented code is recompiled and executed using
randomly generated test cases. If an execution trace matches a trace described
by the threat model, security violations are reported and actions should be
taken to mitigate the threat in the system.

\subsection{Models in model-based security testing}
In order to test security properties information from different
sources are needed and need to be systematically related to each other to support tracing and proper usage of the information. In
addition to the functional system specification and to system architecture
information, information on known or potential vulnerabilities, potential attacks and their occurrence probabilities can
give guidance on what to test and how to test. In addition, probabilities and estimations on the severity of potential
attacks can be summed up to form a risk analysis that points at the threats
that are to be considered. Such a risk analysis provides guidance for test ordering and test
prioritization and supports test management by indicating the need for test recommended test resources. 

Hence, model-based
security testing needs to be based on different types of models in order to cover the
different perspectives used in securing a system. In the
following we provide three different model categories that each represent a
perspective on its own and may serve as input models for
test generation. 

\subsubsection {Architectural and functional models}
Architectural and functional models of the SUT are concerned with
system requirements regarding the general behaviour and setup of a
software-based system. The main perspective of these models are the structure and properties of the system under test. The models exist on different
level of abstraction and in different granularity. Often they show additions that allow to focus on 
specific system properties like robustness
properties (e.g.~failure states) or performance properties (e.g.~durations or throughputs). Regarding security
testing, we are principally interested in locating critical system functionality with respect to the overall software
architecture and in identifying security-critical interfaces, which might be an entry
point for an adversary. Related to security-critical interfaces, interaction models or
protocol models (involving data models or behavioural models) are of high
interest for security testing. In addition
functional security measures (such as authentication or access control means) can be specified
within functional models and be tested by use of functional testing approaches. 

\begin{figure}[htbp]
    \centerline{\includegraphics[width=0.6\textwidth]{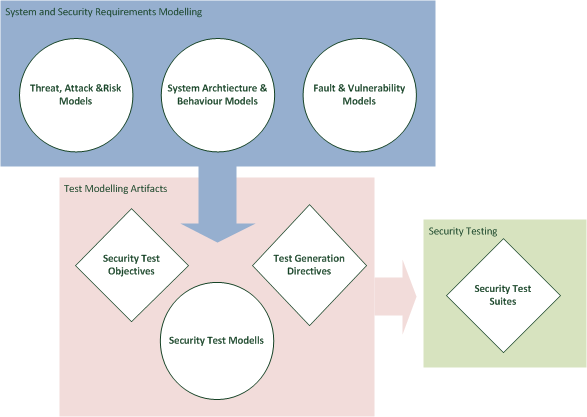}}
    \caption{Modelling Artifacts in MBST}
    \label{fig:TestModels}
\end{figure}

\subsubsection{Threat,  fault and risk models}
While architectural and functional models typically describe the expected system configuration and behaviour, risk modelling techniques like the CORAS risk modelling
approach~\cite{Lund2011,Hogganvik2007} focus on what can go wrong. CORAS provides means to mode risks,
threats or faults and enables the identification of multiple risk factors, describe their relationships
and relate them to occurrence probabilities and potential impacts. 

Besides
CORAS, further approaches for fault and attack modelling exist. Well-known are the fault tree analysis (FTA~\cite{Vesely_Goldberg_Roberts_Haasl_1981}) and the cause-consequence analysis (CCA~\cite{Nielsen1971}). FTA considers high-level
faults and decomposes them top-down to basic events, which can be identified and tested for
in the system. 

A variant of fault trees are so called attack trees~\cite{Mauw05}. Attack trees are directly related to security risks. They start
from a high-level attack scenario and decompose them to
concrete basic interactions with the system. 

ETA (event tree analysis~\cite{reay2002efficient}) works
bottom-up. It starts with the identification of unwanted system events
and analyses the consequences in case of an occurring unwanted event. 

A CCA (cause-consequence analysis~\cite{DCCA-SCADE-SafeComp-07}) -- a combination of the FTA and ETA concepts -- can be used as well. That
analysis starts with a thread. The causes (top-town) and the consequences
(bottom-up) are analyzed simultaneously.

\subsubsection {Weakness and vulnerabilities models} 

While threat, fault and
risk models concentrate on causes and consequences of system failures,
weaknesses or vulnerabilities, a weakness or vulnerability model describes the
weakness or vulnerability by itself. The information needed to develop such
models are normally given by databases like the
National Vulnerability Database (NVD) or the Common Vulnerabilities and
Exposures (CVE) database. These databases collect known vulnerabilities
and provide the information to developers, testers and security experts, so
that they can systematically check their products for known
vulnerabilities. One of the challenges yet not sufficiently solved is how these
vulnerabilities can be integrated in system models, so that they can be used for test 
generation. 

One possible solution is based on the idea of mutation testing~\cite{DBLP:journals/ijsi/WeiglhoferAW09}. Typically, mutation testing is used to qualify test suites by
running tests against a mutation of the system under test. The quality of the
test suite is stated with respect to the number of mutants being detected by the test suite. 
For security testing, models of the system under test are mutated in a way
that the mutants represent weaknesses or known vulnerabilities. These weakness or
vulnerability models can then be used for test generation by various MBT approaches. The generated tests are used to check whether the system under test is weak or vulnerable with respect to the weaknesses and vulnerabilities
in the model.

\subsection{Activities in model-based security testing}

Security testing like any other testing follows a series of activities and uses artifacts that aim to systematically plan, design, specify, realize, and execute tests, and to evaluate the
test results and, if needed, to readjust the planning etc. 

MBST shows slight differences in these
activities, but follows the main sequence of activities. A main task is to provide system properties under consideration concrete test cases (data and behaviour) that represent the stimuli to the system under test and the 
evaluation of the system reactions by test oracles. The following discusses how to generate security tests with a model-based approach.

\begin{itemize}
\item Identify security test objectives and methods: The test objectives
define the overall goals of testing and relate the
goals to testing methods that allow to accomplish the objectives. For model-based testing the modelling techniques and test generation strategies need
to be planned. Especially  \emph{threat, fault, and risk  models} are to be considered
to guide or strengthen the  test identification with respect to the
identified risks, threats, faults, and their consequences.

\item Design a functional test model: The test model reflects either the  expected functional scenarios of the SUT (system perspective) or the scenarios of the SUT usage (system perspective). Standard
modelling languages such as UML can be used to formalize the points of control
and observation of the SUT, the dynamic behaviour when interacting with the system, the
entities associated with the test in various test
configurations, and the test data applied to the system. The test models need to be precise and complete enough to allow
automated derivation of executable tests from these models. However, security testing
focuses either on testing the correctness of security functions or on testing the robustness
against a dedicated misuse of the system. Thus, \emph{functional test models} used for security testing describe not only the typical environment or usage of a system, but also adversary
environments or atypical usages like attacks and hacking attempts.

\item Determine test generation criteria: Usually, there is an infinite
number of possible tests that can be generated from a model, so that test
designers choose test generation or selection criteria to limit the number of (generated) tests to a finite number by e.g. selecting highest-priority tests, or to ensure specific coverage of system structures, behaviours, or alike. For security testing, approaches
based on structural model coverage, i.e. determining the coverage of model
elements by generated tests, is not sufficient. Hence, fuzz test approaches are
used that follow other kind of coverage criteria and lead to different, but also larger number
of generated test data or test behaviour. However, approved test generation
criteria like the coverage of security functional requirements or the weighting
of security functional requirements with risk values are applicable as well.

\item Generate the tests: The test generation is in MBT typically a fully
automated process to derive the test cases from a given test model as determined by the chosen test generation criteria. This is also true for MBST. The generated test cases are sequences of high-level events or actions to or from the SUT, with input
parameters and expected output parameters and return values for each event or
action. If needed, the generated tests are further refined to a more concrete
level or adapted to the SUT to support their automated execution.

\item{Assess the test results: During test result evaluation and  test assessment, the
quality of the SUT can be rated with respect to the test results as well as the quality of 
tests can be rated with respect to their fault and vulnerability revealing capabilities. While for
functional MBT, measurements and metrics for system and test quality exist this is still a research challenge for MBST. Furthermore, the
test results need to be analyzed if changes to the system
requirements, the system design, the risk analysis or to the test process itself are needed. In these cases, required iterations are to be started.}
\end{itemize}

\subsection{Testing Approaches in DIAMONDS}

The project DIAMONDS (Development and Industrial Application of Multi-Domain
Security Testing Technologies) develops under the direction of Fraunhofer FOKUS,
Berlin, efficient and automated security test methods for security-critical,
networked systems in various industrial domains such as industrial automation,
banking and telecommunications. DIAMONDS develops methods to design objective,
transparent, repeatable, and automated security tests that focus on system
specifications and related risks. The project goals include the definition of security fault and
vulnerability modelling techniques,
the definition of a security test pattern catalogue, the development of MBST techniques, and the definition of a MBST methodology. DIAMONDS examines vulnerabilities of networked systems in
the considered domains in order to derive common principles, methods and means
that enable effective security testing of industrial importance. In reflection
of the case studies results, the DIAMONDS security testing methodology will
be evaluated and optimized. The project results are made available to interested parties and also through contributions to the standardization at ETSI and to other standardization bodies. 

A special focus is given in DIAMONDS to (1) risk-based MBST and to (2) model-based fuzz testing.

\subsubsection{Risk-based security testing}

Risk-based testing can be generally introduced with two different goals
in mind. On the one hand side risk-based testing approaches can help to optimize the overall test process. The results of the risk analysis, i.e. the results of threat and vulnerability analysis, are used to guide the test identification and may complement requirements engineering results with systematic information concerning threats and vulnerabilities of a system. 
On the other hand side, attack simulation is to find deviations of the SUT to its specification that leads to vulnerabilities because invalid inputs are not rejected but processed by the SUT instead. Such deviations may lead to undefined states of the SUT and can be 
exploited by an attacker, for example to successfully perform a
denial-of-service.

A comprehensive risk
assessment additionally introduces the notion of risk values, that is the
estimation of probabilities and consequences for certain threat scenarios. These
risk values can be additionally used to weight threat scenarios and thus help
identifying which threat scenarios are more relevant and thus identifying the
threat scenarios that are the ones that need to be treated and tested more
carefully. 

Furthermore, risk-based testing approaches can help to optimize the
risk analysis and the risk assessment itself. Risk analysis and risk assessment,
similar to other development activities in early project phases, are mainly
based on assumptions on the system itself. Testing is one of
the most relevant means to do real experiments with a system and thus enables to
gain empirical evidence on the existence of vulnerabilities, the applicability
and consequences of threat scenarios and the quality of countermeasures. Thus,
risk-based testing results can be used as a form of evidence for the assumptions
that have been made during the risk evaluation and risk assessment. 

In
particular, risk-based testing can help in

\begin{itemize}
\item providing evidence on the functional correctness of countermeasures,
\item providing evidence on the absence of known vulnerabilities, and
\item discovering unknown vulnerabilities,
\item optimizing risk analysis
by identifying new risk factors and reassessing the risk values.
\end{itemize}

The CORAS language~\cite{Hogganvik2007} integrates different tree-based approaches for risk
modelling. It is a graph-based modelling approach that emphasizes the
 modelling of threat scenarios and provides
formalisms to annotate the threat scenarios with probability values and
formalisms to reason with these annotations. 
 
\begin{figure}[htbp]
    \centerline{\includegraphics[width=0.75\textwidth]{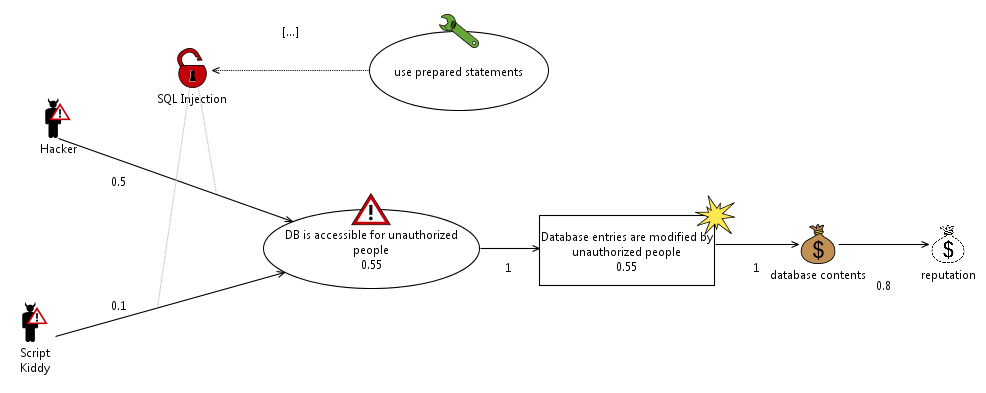}}
    \caption{CORAS treatment diagram}
    \label{fig:coras_threat}
\end{figure}

Figure~\ref{fig:coras_threat} shows a simple CORAS risk model that depicts a threat scenario for an unauthorized database access. The CORAS language allows to relate threat scenarios to adversaries (so called threats, e.g. "`hacker"' or "`script kiddy"' in Figure~\ref{fig:coras_threat}) and to potential vulnerabilities. A vulnerability is denoted by the
unlocked padlock (see e.g. "`SQL injection"'). Last but
not least the threat scenario is related to unwanted incidents, e.g. "`Database
entries are modified by unauthorized people"'. This modification harms the asset "`database
contents"' (see "`brown moneybag"' in Figure~\ref{fig:coras_threat}) and can therefore negatively influence the asset "`revenue"' (see "`white moneybag"'). The treatment
"`use of prepared statements"' instead of strings containing SQL queries is depicted with the
green pliers in that figure. Except the threats, the
vulnerability, the assets and the treatment, all elements have annotations that
denote the probability or frequency of the transition or the incidence of for instance a threat scenario. 

Such risk models can be used in different ways to support testing. The goal of \emph{risk-based test identification }is to improve the
test design such that high-risk areas of the SUT are covered and that at the same time test resources are optimally used by focusing on highest risks first. 

In risk-based test identification, for individual risk factors test objectives are developed. We
consider a test objective to be foremost an informal specification that defines
which aspect of a certain system, functionality, or protocol etc. should be
tested. Similar to requirements in requirements engineering, test
objectives constitute test requirements that can be refined and decomposed during the test development. In
the following we describe the relationship between test objectives
and the elements used in the risk analysis. Quantifications of the related risks can be used to
weight the test objectives.

Test objectives for
\begin{itemize}
\item an unwanted incident describe which test
methods can be applied to initiate and detect an unwanted incident and to
characterize its consequences. 

\item threat scenarios describe which test methods
can be applied to initiate a threat scenario and to characterize its
consequences. 

\item vulnerabilities describe which test methods
can be applied to elicit a vulnerability. 

\item treatment scenarios describe which test methods can be applied to characterize the maturity and effectiveness of a
treatment scenario. 
\end{itemize}

Another way of using risk models in testing is \emph{risk-based test selection}. It is used to find an optimal set of test cases along certain selection strategy. The selection strategy takes into
account available test resources and optimizes the selected tests with respect to the chosen coverage criteria. In functional testing coverage is often
described by the coverage of requirements or by the coverage of system or test model elements such
as states, transitions, or decisions. In risk-based testing, we aim at the
coverage of system risks. The
criteria are designed by taking the risk values from the risk assessment to set
priorities for the test generation or to order the test
execution in a test run. The test selection can be either accomplished on existing test cases to select tests for test run or during test generation to enable the directed, goal-oriented generation of tests.

Last but not least, security testing supports \emph{risk control}. 
Risk control deals with the revision of risk assessment results by correcting
assumptions on probabilities, consequences or the maturity of treatments
scenarios or deals with the completion of risk analysis result by integrating
vulnerabilities and thus potentially threats, threat scenarios and unwanted
incidents. The test
results can reflect and verify the assumptions that have been made during risk
analysis. The test results can be used to adjust risk analysis
results by introducing new or revised vulnerabilities or revised risk
estimations on basis of the defects being found. Test results,
test coverage information and a revised or affirmed risk assessment can provide
 solid arguments for the security level of
a system as test results relate to the

\begin{itemize}
\item risks, which are addressed and covered by related
test cases.

\item treatment or threat scenarios, which are to be checked by the test cases.

\item assets, whose related risks are checked by corresponding test cases.

\item vulnerabilities, which have been identified and located related test cases.
\end{itemize}

\subsubsection{Model-Based Fuzzing}

While the origin of fuzzing is based on a complete
randomized approach~\cite{Miller90anempirical},
block-based and model-based fuzzers use their knowledge about the message
structure to systematically generate messages containing invalid data among
valid data~\cite{takanen2008fuzzing}.

Systematic approaches are often more successful because the
message structure is preserved and thus the likelihood increases that the
generated message is accepted by the SUT. Using fuzz testing principles not only 
for test data generation but also for test behaviour generation complements
the traditional fuzz testing approaches. Behaviour fuzzing does not only reflect the 
generation of atypical messages but also changes the typical appearance and
order of messages. For example a valid and approved sequence of messages can be
turned into an atypical and unknown sequence by rearranging messages, repeating
and dropping them or just by changing the type of message.

Behaviour fuzzing aims at finding flaws in design and vulnerabilities in
systems that are not simply revealed by applying invalid input data. It
focuses on misuse on a higher level of functionality. For example, a security
requirement defines that a download may only be started after successful
authentication. In a vulnerable system the download can be started additionally without any
authentication. Such fault can be detected using typical input data but atypical
behaviour, e.g. by simply omitting the authentication.

\begin{figure}[htbp]
    \centerline{\includegraphics[width=0.25\textwidth]{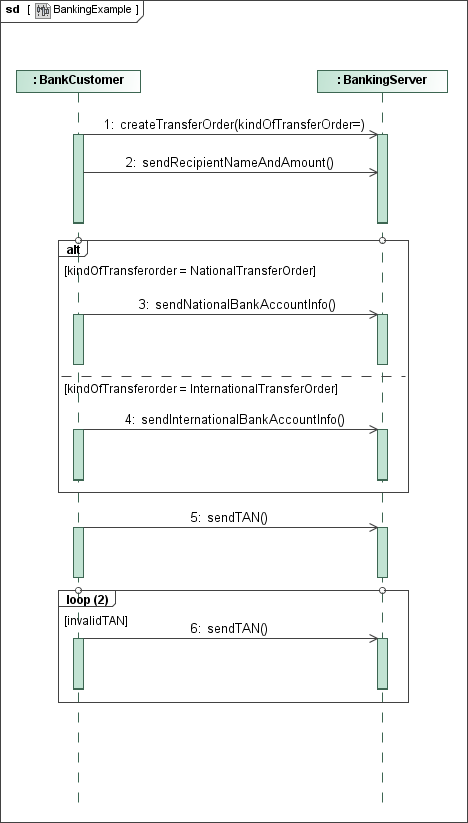}}
    \caption{Transfer Order Sequence}
    \label{fig:seq_transfer}
\end{figure}

DIAMONDS develops model-based fuzzing approaches that use e.g. fuzzing operators on scenario models which are specified by sequence diagrams.  
In the following, a simplified example from the banking domain is used to
illustrate how fuzzing operators are applied to sequence diagrams. For ease of
understanding most parameters are omitted.

The sequence diagram in Figure~\ref{fig:seq_transfer} describes how a bank
customer can perform a transfer order. The customers can either order a national
or an international transfer (message 1). Afterwards the customer sends the name
of the recipient of the transfer order, the amount to be transferred (message
2) as well as recipient's national bank account information (message 3) in
case of a national transfer order or recipient's international bank account
information (message 4) in case of an international transfer order. The transfer
order must be authorized by the customer sending a valid transaction number TAN
(message 5). If the customer accidentally sent an invalid TAN e. g. by mistyping
it, he can try to enter a valid TAN up to two times again (message 7, combined
fragment loop).

\begin{figure}[htbp]
    \centerline{\includegraphics[width=0.75\textwidth]{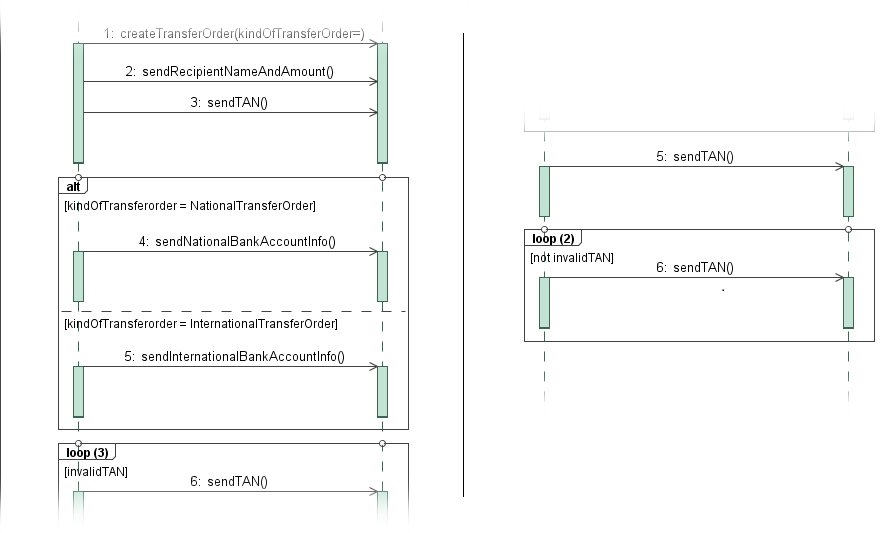}}
    \caption{Fuzzed Transfer Order Sequence}
    \label{fig:seq_transfer_fuzz}
\end{figure}

Applying fuzzing operators to the diagram, messages can be moved, removed,
repeated, inserted or the type of a message can be changed to obtain an invalid
sequence. Fuzzing operators perform a mutation on the diagram resulting in an invalid sequence
in comparison with the original. One fuzzing operators performs only one mutation of a sequence
diagram. For instance, a fuzzing operator can move message 5 after message 2. Another
fuzzing operator can generate an invalid sequence of messages by negating interaction constraints
of interaction operands. By negating the interaction constraint of the loop
combined fragment, the sequences generated from the resulting sequence diagram
contain at least two valid transaction numbers sent to the banking
server (three if the second given TAN is valid). Figure \ref{fig:seq_transfer_fuzz} shows
the results of the fuzzing operators from above.

Performing the above mentioned fuzzing operators leads to different sequence
diagrams that are the basis for further test case generation. However, the main idea
of fuzz testing approaches in general is the ability to automatically generate a
large number of test cases. This is achieved by applying not only one fuzzing operator
to a sequence diagram, but a set of fuzzing operators multiple times, e.g. by applying
a single fuzzing operator to several model elements of a
sequence diagram or by applying several, possibly different fuzzing operators, one after another. 
The combination of fuzzing operators permits the generation of a large number of test cases.

\section{Summary}

Model-based security testing (MBST) is a relatively new field and especially dedicated
to the systematic and efficient specification and documentation of security test
objectives, security test cases and test suites, as well as to their automated or
semi-automated generation. This paper provides an initial survey on model-based security testing by analyzing related work, discussing models that can be used for model-based security testing, and by outlining two main approaches that are being developed in the European ITEA2 project DIAMONDS by industrial and research partners from 6 countries:

\begin{itemize}
	\item Risk-based security testing
	\item Model-based fuzzing
\end{itemize}

Details of risk-based security testing and model-based fuzzing are given in the DIAMONDS deliverables.  While DIAMONDS is an ongoing project that is at the half of the project duration having reached 2 of 4 milestone, the methods are still under development and the analysis of the gains and the pros and cons of the methods is still to be done. However, initial versions of the methods have already been applied in selected case studies that demonstrated the potentials of the described approaches.

\bibliographystyle{eptcs}
\bibliography{MBSecTest}
\end{document}